 \documentclass[smallabstract,smallcaptions]{dccpaper}

\usepackage[utf8]{inputenc} 
\usepackage[T1]{fontenc}    
\usepackage{citesort}
\usepackage{epsfig}
\usepackage{color}
\usepackage{cite}
\usepackage[pagebackref=false,breaklinks=true,colorlinks,citecolor=black,linkcolor=black,urlcolor=black,bookmarks=false]{hyperref} 
\usepackage{url}   
\usepackage{float}
\usepackage{graphicx}
\usepackage{subfig}
\usepackage{amsmath}
\usepackage{array}
\usepackage{multirow}
\usepackage{flushend}
\usepackage{amssymb}
\usepackage{amsfonts}

\newlength{\figurewidth}
\newlength{\smallfigurewidth}

\begin{document}

\title
{\large
\textbf{Lossless Point Cloud Attribute Compression Using Cross-scale, Cross-group, and Cross-color Prediction}
}

\author{%
Jianqiang Wang$^\dag$, Dandan Ding$^\ddag$, and Zhan Ma$^\dag$\\
$^\dag$Nanjing University, $^\ddag$Hangzhou Normal University\thanks{This work was partially supported by the National Natural Science
Foundation of China under Grant 62022038 and Grant U20A20184. (Corresponding Author: Zhan Ma (mazhan@nju.edu.cn)).}
}

\maketitle
\thispagestyle{empty}
\vspace{0.2cm}
\begin{abstract}
This work extends the multiscale structure originally developed for point cloud geometry compression to point cloud attribute compression. To losslessly encode the attribute while maintaining a low bitrate, accurate probability prediction is critical. With this aim, we extensively exploit cross-scale, cross-group, and cross-color correlations of point cloud attribute to ensure accurate probability estimation and thus high coding efficiency. Specifically, we first generate multiscale attribute tensors through average pooling, by which, for any two consecutive scales, the decoded lower-scale attribute can be used to estimate the attribute probability in the current scale in one shot.  Additionally, in each scale, we perform the probability estimation group-wisely following a predefined grouping pattern. In this way, both cross-scale and (same-scale) cross-group correlations are exploited jointly. Furthermore, cross-color redundancy is removed by allowing inter-color processing for YCoCg/RGB alike multi-channel attributes.  The proposed method not only demonstrates state-of-the-art compression efficiency with significant performance gains over the latest G-PCC on various contents but also sustains low complexity with affordable encoding and decoding runtime.

\end{abstract}
\section{Introduction}
Point cloud uses a great number of unconstrained 3D points to render 3D objects and scenes realistically, in which each point is formed using the coordinate $(x,y,z)$ and associated attributes such as the color (RGB or YCoCg), normal, and reflectance. Nowadays, point clouds have been massively used in networked applications including Augmented and Virtual Reality, Autonomous Machinery, etc., making the desire for efficient Point Cloud Compression (PCC) more and more indispensable. In addition to those rules-based PCC solutions, such as Geometry-based PCC (G-PCC) or Video-based PCC (V-PCC) standardized under the ISO/IEC MPEG committee~\cite{cao2021compression}, learning-based PCC approaches have attracted worldwide attention and demonstrated noticeable compression gains~\cite{quach2022survey} in Point Cloud Geometry Compression (PCGC). Among them, our earlier multiscale sparse representation-based PCGC has reported state-of-the-art performance~\cite{Wang2021SparseTM,Wang2021MultiscalePC} on a variety of point clouds (e.g., dense object and sparse LiDAR data). This work, furthermore, extends the multiscale approach for lossless Point Cloud Attribute Compression (PCAC). Following the convention~\cite{MPEG_GPCC_CTC}, losslessly compressed geometry is assumed to study the PCAC.

Similar to the geometry occupancy compression, the attribute compression performance also depends on the accuracy of probability estimation~\cite{Wang2021SparseTM}. However, the dynamic range of the Point Cloud Attribute (PCA) is generally much wider than the binary occupancy status (e.g., 1 for occupied and 0 for non-occupied), e.g., as for the color attributes, the intensity of an 8-bit RGB sample is typically ranging from 0 to 255. Thus, a better solution is highly desired to exploit redundancy across attributes for compact representation. With this aim, we leverage the cross-scale, cross-group, and cross-color\footnote{Cross-color prediction is feasible for multi-channel attributes like three-channel RGB/YCoCg colors.} predictions, which thoroughly examine the point correlation to improve the attribute probability approximation in compression.

\begin{figure*}[t]
\centering
\includegraphics[width=6in]{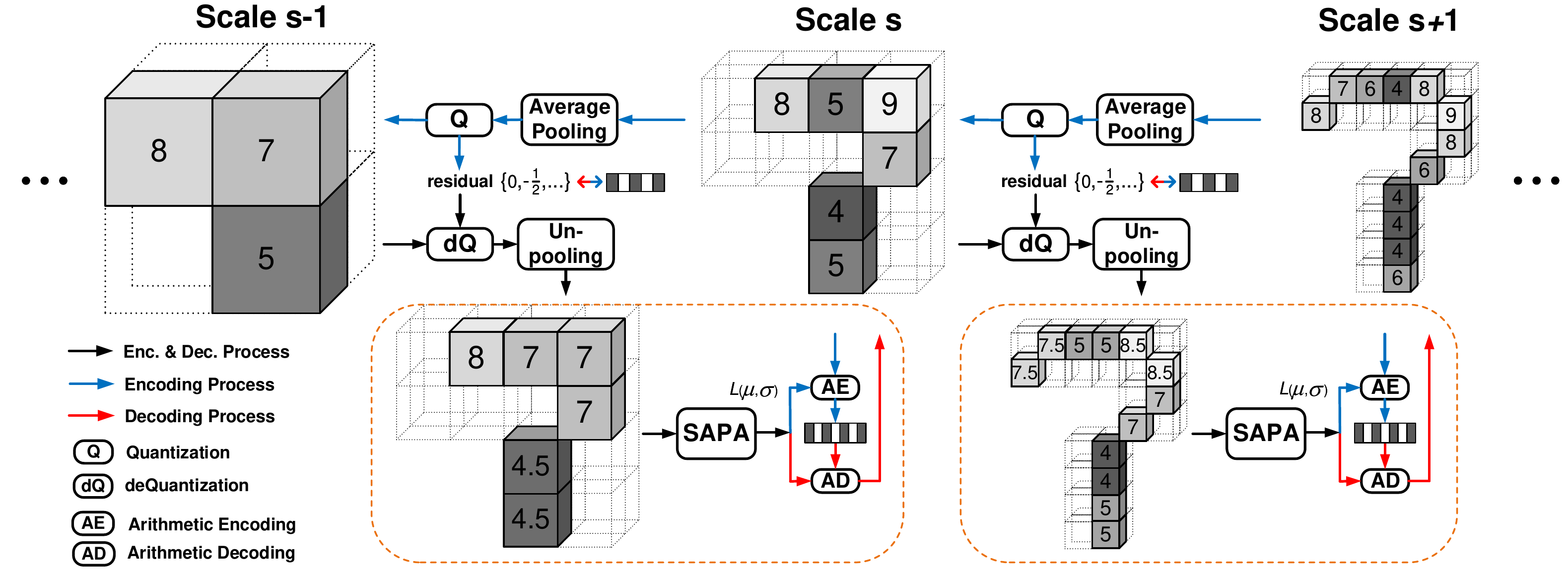}
\caption{{\bf Cross-scale Prediction.} Average pooling is used in encoding to progressively generate multiscale representations. The decoded/processed lower-scale attributes are used to approximate the attribute probability of the current scale. Rounding-based quantization is used. The scale-wise {quantization residual} is randomly distributed and compressed assuming the {uniform} distribution.  The SAPA leverages SparseCNN to effectively characterize neighborhood correlation for probability estimation.}
\label{fig:multiscale}
\end{figure*}

As shown in Fig.~\ref{fig:multiscale}, {for the encoding process, the original point cloud tensor is downscaled progressively to form multiscale tensors, where the geometry is scaled dyadically and the associated attributes are averagely pooled; while the decoding mirrors the encoding steps to upscale the geometric tensor and unpool the attributes accordingly.}
Upon such multiscale representation, cross-scale prediction can be applied, where decoded lower-scale attributes are leveraged to estimate the probability of current-scale attributes for encoding/decoding. Currently, we enforce the integer attribute at each scale through a simple rounding-based quantization. And random quantization residual is compressed assuming the uniform distribution.

Instead of estimating the attribute probability of all valid points at the current scale in one shot using lower-scale information only, we can optionally approximate the probability from one group to another following a predefined grouping pattern shown in Fig.~\ref{fig:multistage}, by which we best utilize the lower-scale and inter-group correlations (same-scale) jointly.  Parallel processing is applied for intra-group computation.

Furthermore, as there exist inter-color correlations even after the color space conversion~\cite{CCP_HEVC}, we can further reduce the compression bitrate by enforcing the cross-color prediction. Here, we use the YCoCg color space as the standardized G-PCC~\cite{GPCC}.  The compression at a given scale follows the processing order of Y, Co, and Cg, where the processing of Co (Cg) can use the previously-processed Y (Y and Co), as in Fig.~\ref{fig:multichannel}.  Note that for the attribute with a single-channel component, e.g., the reflectance in sparse LiDAR point clouds, we only use cross-scale and cross-group prediction.

\begin{figure*}[b]
\centering
\includegraphics[width=5.5in]{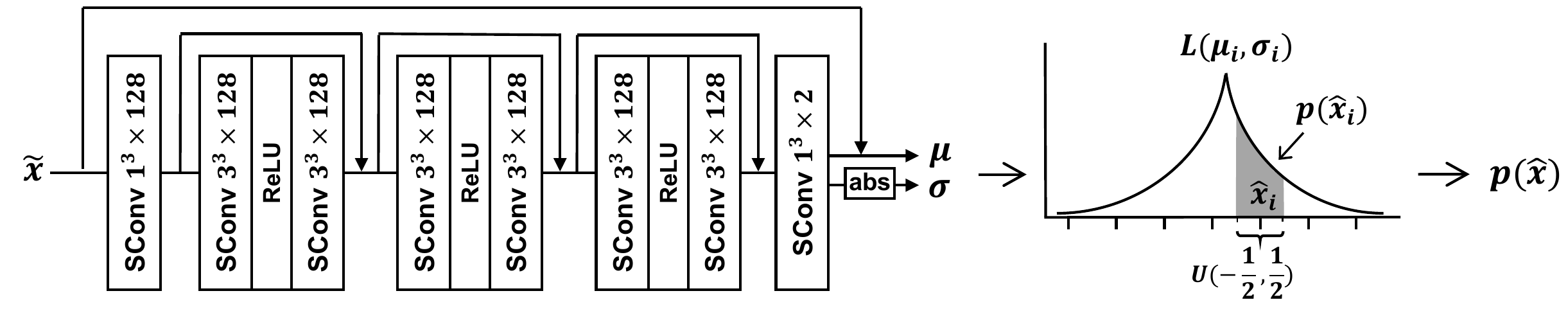}
\caption{{\bf SAPA.} SparseCNN-based Attribute Probability Approximation.}
\label{fig:network}
\end{figure*}

The above attribute probability approximation using cross-scale, cross-group, or cross-color prediction is facilitated using Sparse Convolutional Neural Networks (SparseCNNs), which is referred to as the SAPA in Fig.~\ref{fig:network}. The SAPA model inputs valid neighbors to produce associated Laplacian distribution parameters for the probability derivation. The advantages of sparse convolutions can be found in~\cite{choy20194d}.

{\bf Contributions.}
1)  To the best of our knowledge, this work is probably {\it the first} lightweight and generalized lossless point cloud attribute compression approach that outperforms the latest G-PCC, e.g., about 15\%, 12\%, 37\%, 10\%, and 5\% bitrate reduction for respective 8iVFB, Owlii, MVUB, ScanNet and Ford datasets;  2) The outstanding compression performance comes with the elegant utilization of neighborhood correlations through the cross-scale, cross-group, and cross-color prediction for attribute probability estimation; 3) The lightweight computation is due to the use of sparse convolution (to reflect the sparsity of point clouds) and parallel processing inherently supported by our design, demonstrating similar encoding/decoding runtime as the G-PCC codec.

\section{Related Works} \label{sec:related_work}

\textbf{Point Cloud Attribute Compression (PCAC).}

{\it Rules-based solutions} mostly utilize transforms to exploit attribute correlations across neighborhood points. For example, Graph Fourier Transform (GFT) and its variants were examined in~\cite{zhang2014point,cohen2016attribute,shao2017attribute,Xu2021PredictiveGG}. Similarly, Gaussian Process Transforms (GPTs) were also studied in~\cite{de2017transform}. 
Queiroz~\textit{et al.}~\cite{de2016compression} proposed a Region-Adaptive Hierarchical Transform (RAHT), a hierarchical sub-band transform that resembles an adaptive variation of a Haar wavelet, which was adopted in the MPEG G-PCC. The latest G-PCC version (TMC13v14), notably improved the original RAHT with state-of-the-art PCAC efficiency. G-PCC also adopted the Predicting and Lifting Transforms to compress the attribute according to the organization of the Level of Details.

{\it Learning-based approaches} have attracted extensive attention recently.
Quach~\textit{et al.}~\cite{quach2020folding} folded 3D point cloud attributes onto the 2D grids and then directly applied the conventional 2D image codec for compression.  
Fang~\textit{et al.}~\cite{fang20223dac} designed an MLP-based entropy model to approximate the probability of RAHT coefficients. 
Alternatively, end-to-end PCAC was also studied.  
Deep-PCAC~\cite{sheng2021deep} applied a point-based network to compress point cloud attribute, while Alexiou~\textit{et al.}~\cite{alexiou2020towards} directly used 3D dense convolutions for compression. 
Recently, SparsePCAC~\cite{Wang2022SparsePCAC} was developed to process the sparse tensor under the variational autoencoder structure for efficient attribute representation.
Unfortunately, despite the technical progress provided by these learning-based PCAC solutions, the lossy compression efficiency is still inferior to the latest G-PCC, not to mention that some of them are exceptionally complex for practical application~\cite{quach2020folding,sheng2021deep,alexiou2020towards}.

\textbf{Multiscale Sparse Representation based PCGC.} Recently, the compression of point cloud geometry has been significantly improved by applying learning-based solutions to effectively exploit correlations.  Among them, multiscale sparse representation-based solutions have demonstrated leading compression performance in both lossy and lossless modes for a variety of point clouds~\cite{Wang2021SparseTM}. This work extends the multiscale structure to support PCAC by exhaustively exploiting cross-scale, cross-group, and cross-color (if applicable) correlations.

\section{Method}
Let {\{$x_i^{(S)},i=1,2,\ldots, N^{(S)}$\}} represent the original PCA tensor at the highest scale $S$. $N^{(S)}$ is the total number of positively-occupied voxels (POV) at scale $S$ and
$x_i^{(S)}$ is the attribute intensity for $i$-th POV which is an integer in general. Here single-channel attributes are exemplified with the intensities annotated  in Fig.~\ref{fig:multiscale}.

\subsection{Cross-scale Prediction}

\textbf{Average Pooling \& Quantization.}
As in Fig.~\ref{fig:multiscale}, for the {$s$-th} scale PCA tensor \{$\hat{x}_i^{(s)}, i=1,2,\ldots, N^{(s)}$\}, $k$ valid POVs in a $2\times2\times2$ cube are merged into one POV at $(s-1)$-th scale with its attribute value
    $x^{(s-1)} = \frac{1}{k}\sum_{m=1}^k\hat{x}_{m}^{(s)}$,
 having ${k \in \{1,2,\cdots,8\}}$. 
$x^{(s-1)}$ is then rounded to an integer, e.g., $\hat{x}^{(s-1)}=\lfloor x^{(s-1)} \rceil$, producing the residual $r^{(s-1)}=x^{(s-1)}-\hat{x}^{(s-1)} \in \{0,-\frac{1}{2},\cdots\}$ and the $(s-1)$-th scale tensor \{$\hat{x}_j^{(s-1)}$\}. The same pooling and quantization process will be repeated until the arrival of the lowest-scale tensor that only contains one POV as the global average of the input tensor, e.g., \{$\hat{x}_{j=1}^{(s=1)}$\}.  As a result, the PCAC problem is reformulated as the compression of 
{\{$\hat{x}_i^{(s)}$\} and \{${r}_j^{(s-1)}$\} with $s\in[1, S]$,  $i\in[1, N^{(s)}]$ and $j\in[1, N^{(s-1)}]$},
where the key challenge is to estimate an accurate probability for each element to be coded.

For the compression of quantized residual \{${r}_j^{(s-1)}$\}, we directly encode them under the uniform distribution, because it is randomly distributed among values 0, $-\frac{1}{k}$ and $\frac{1}{k}$. Here $k$ is the number of valid POVs in a local $2\times2\times2$ cube.
Next, to encode/decode each element in \{$\hat{x}_i^{(s)}$\},  $i\in[1, N^{(s)}]$,  we perform the probability estimation by using decoded lower-scale attributes \{$\hat{x}_j^{(s-1)}$\}, $j\in[1, N^{(s-1)}]$. More specifically, \{$\hat{x}_j^{(s-1)}$\} is first augmented with decoded quantization residual \{$r_j^{(s-1)}$\} to recover  \{$x_j^{(s-1)}=\hat{x}_j^{(s-1)}+r_j^{(s-1)}$\}.
Each element in \{$x_j^{(s-1)}$\} is associated with a corresponding $j$-th POV, which is then dyadically upscaled to eight child nodes as a local $2\times$2$\times2$ cube $j$. Since the geometry occupancy is available in advance, all $k$ valid POVs in this cube $j$ are filled  with the same attribute intensity $x_j^{(s-1)}$, e.g., 
{$\{\tilde{x}_{m}^{(s)}\}_{j} = x_j^{(s-1)}$, $m\in[1, k]$; $\{\tilde{x}_{m}^{(s)}\}_{j}$ contains $k$ POVs in $j$-th $2\times$2$\times2$ cube}
\footnote{Normally $k<8$ because of the sparsity nature of the point clouds.}. Apparently, it is the unpooling process that  produces the upscaled tensor \{$\tilde{x}_i^{(s)}$\}.  Subsequently, this upscaled tensor \{$\tilde{x}_i^{(s)}$\} is fed into the SAPA to produce element-wise probability  for \{$\hat{x}_i^{(s)}$\}. This scale-wise processing is repeated across all scales.

\textbf{SAPA.} Figure~\ref{fig:network} shows the SAPA model which consists of multiple sparse convolutional layers (SConv), nonlinear activation layers (ReLU), and residual links. Let us take the process from the scale $s-1$ to  $s$ as an example. The input to the SAPA model is \{$\tilde{x}_i^{(s)}$\}, and the outputs are respective mean \{$\mu_i$\} and variance \{$\sigma_i$\} for each element in Laplacian distributed \{$\hat{x}_i^{(s)}$\}. As a result, the probability  can be computed by integrating the Laplacian distribution $\mathcal{L}$ via 
\begin{align}
    p(\{\hat{x}_i^{(s)}\}) 
    =\prod_i\nolimits\Bigl(\mathcal{L}(\mu_{i}, \sigma_{i})*\mathcal{U}(-\tfrac{1}{2},\tfrac{1}{2})\Bigr)(\hat{x}_{i}^{(s)})
    \quad
    \text{with~}\mu_i, \sigma_i=\text{SAPA}(\{\tilde{x}_i^{(s)}\}),
\label{eq:entropy}
\end{align} 
where $\mathcal{U}(-\tfrac{1}{2},\tfrac{1}{2})$ is the uniform distribution ranging from $-\tfrac{1}{2}$ to $\tfrac{1}{2}$. 
Finally, with the probability, we use arithmetic coding to losslessly encode the attribute intensity into the bitstream or decode it from the bitstream. The corresponding bits can be approximated using $R^{(s)} = \sum\nolimits_i-\log_2(p(\hat{x}_{i}^{(s)}))$.

\subsection{Cross-group Prediction}

\begin{figure*}[t]
\centering
\includegraphics[width=5.6in]{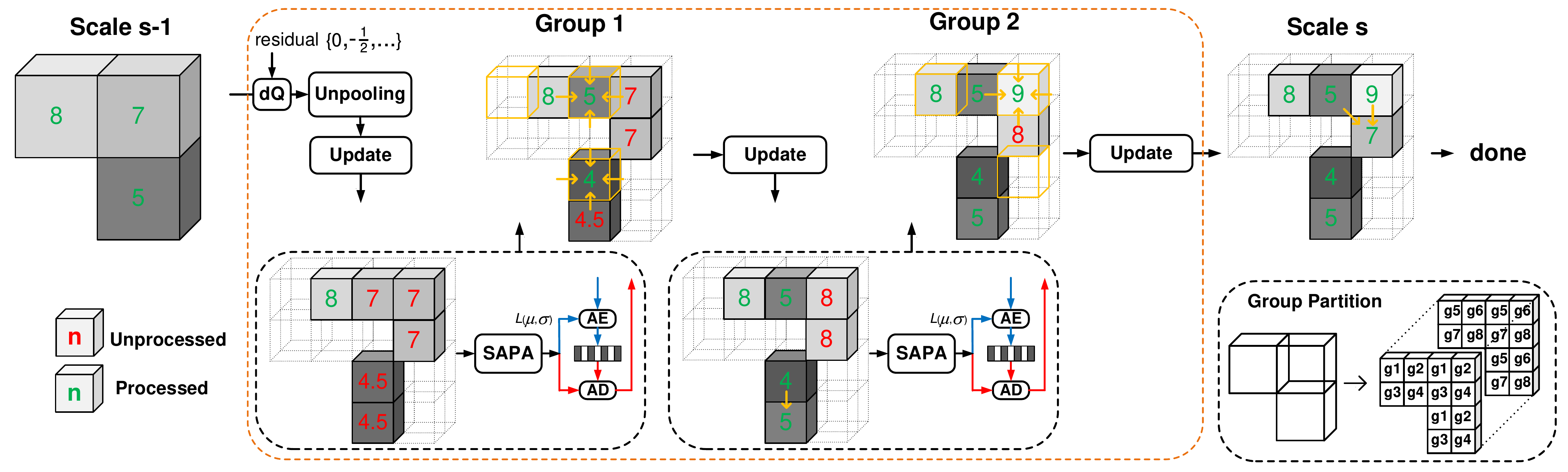}
\caption{{\bf Cross-group Prediction.} Group-wise processing improves the probability estimation through the utilization of lower-scale and inter-group correlations while still maintaining high-throughput computation due to intra-group parallel processing.}
\label{fig:multistage}
\end{figure*}

\begin{figure*}[b]
\centering
\includegraphics[width=4in]{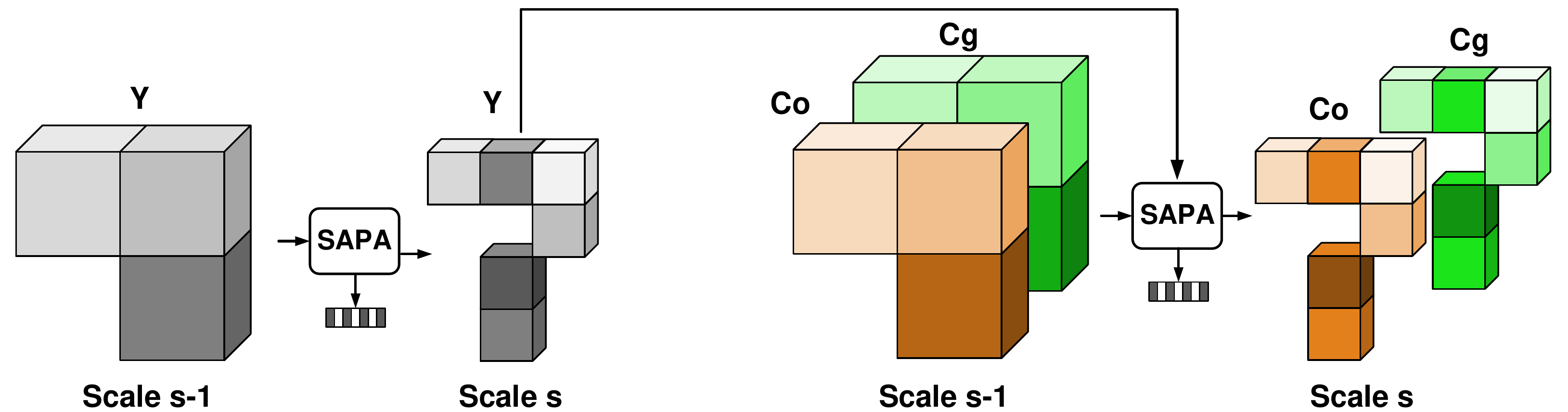}
\caption{{\bf Cross-color Prediction.} Color-wise processing further exploits cross-color redundancy. YCoCg color space is used following the same rule in G-PCC~\cite{GPCC}.}
\label{fig:multichannel}
\end{figure*}

In addition to cross-scale prediction, the accuracy of probability estimation can be further improved by implementing cross-group prediction at the same scale.
As in Fig.~\ref{fig:multistage}, after the dequantization (dQ) and unpooling operations, each POV in the scale $(s-1)$ is expanded to a local 2$\times$2$\times$2 cube with eight child nodes which are partitioned into eight groups {$\{\hat{x}^{(s,g)}, g=1,2,\cdots,8\}$,} according to the geometric positions depicted in the bottom-right part of the plot. Since we have the full knowledge of geometry occupancy,  $k$ valid POVs in this local cube are properly filled with attribute intensities from {\{$\tilde{x}_i^{(s)}$\}}, which are unpooled from the lower-scale as mentioned above.

We then  group-wisely process the probability prediction and encode/decode associated attributes of corresponding POVs from one group to another. For those POV elements in the same group {$\{\hat{x}_{i}^{(s,g)}\}$}, parallel computations are applied.
The SAPA model mentioned above inputs the attribute of valid neighbors in close proximity, e.g., some are already decoded (or inferred) and some are initiated using \{$\tilde{x}_i^{(s)}$\} upscaled from the decoded lower-scale attributes, to derive more accurate probability for the attribute compression of upcoming POVs. 

An additional Update step is introduced at each stage to correct the attribute of those unprocessed POVs initiated using  \{$\tilde{x}_i^{(s)}$\} with the help of previously-processed POVs in proceeding groups, i.e., 
\begin{equation}
    \text{U}(\hat{x}_{i}^{(s,m)}) = \frac{k \times x^{(s-1)} - \sum (\hat{x}^{(s,1)}, \hat{x}^{(s,2)},\cdots,\hat{x}^{(s,g)})}{k-g}
    \quad
    \text{with~} g < m \leq 8.
    \label{eq:update}
\end{equation}
where $\{\hat{x}^{(s,1)},\hat{x}^{(s,2)},\cdots,\hat{x}^{(s,g)}\}$ are processed groups, $\{\hat{x}_{i}^{(s,m)}, m=g+1, \cdots, 8\}$ are unprocessed groups, 
$x^{(s-1)}$ are the average values of these groups generated by $2\times2\times2$ average pooling aforementioned, and $k$ is the number of POVs in each $2\times2\times2$ cube.
This Update process initializes a more accurate attribute value for subsequent computations.
When there is only one unprocessed POV left in each $2\times2\times2$ cube, we can directly infer its ground-truth value without any signaling. 
In total, we only need to compress $N^{(s)} - N^{(s-1)}$ attribute values of POVs at the scale $s$, further saving the compression bitrate.

\subsection{Cross-color Prediction}

Inspired by inter-color correlations studied in~\cite{CCP_HEVC}, we propose the cross-color prediction that first processes the Y channel component, then the Co with the help of Y and Cg with the help of both Y and Co. To accelerate the prediction speed, we process Co and Cg components in parallel in our implementation as shown in Fig.~\ref{fig:multichannel}.

\section{Experimental Results}\label{sec:exp}

\begin{figure*}[t]
\centering
\includegraphics[width=5.5in]{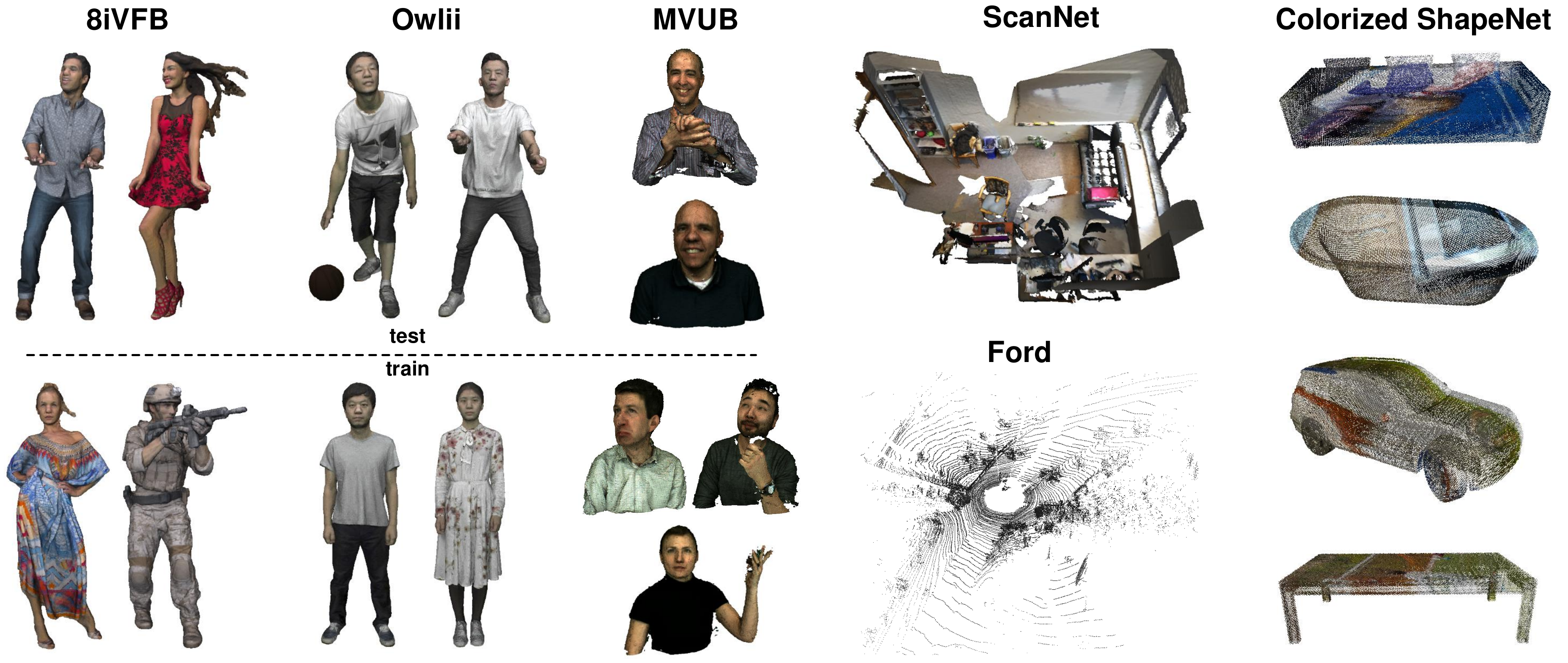}
\caption{{Datasets Used In This Study.}}
\label{fig:dataset}
\end{figure*}

\begin{table}[t]\footnotesize
\centering
\caption{Evaluation of compression efficiency and computational complexity. }
\label{table:main}
\begin{tabular}{|cccccccccc|}
\hline
\multicolumn{2}{|c|}{\multirow{2}{*}{\textbf{PCs}}}                                                                                           & \multicolumn{2}{c|}{\textbf{G-PCC}} & \multicolumn{2}{c|}{\textbf{\begin{tabular}[c]{@{}c@{}}Ours\\ CS\end{tabular}}} & \multicolumn{2}{c|}{\textbf{\begin{tabular}[c]{@{}c@{}}Ours\\ CS+CG\end{tabular}}} & \multicolumn{2}{c|}{\textbf{\begin{tabular}[c]{@{}c@{}}Ours\\ CS+CG+CC\end{tabular}}} \\ \cline{3-10} 
\multicolumn{2}{|c|}{}                                                                                                                        & \multicolumn{2}{c|}{\textbf{bpp}}   & \textbf{bpp}                & \multicolumn{1}{c|}{\textbf{gain}}                & \textbf{bpp}                & \multicolumn{1}{c|}{\textbf{gain}}                   & \textbf{bpp}                            & \textbf{gain}                               \\ \hline
\multicolumn{1}{|c|}{\multirow{3}{*}{\textbf{\begin{tabular}[c]{@{}c@{}}8iVFB\\ vox10\end{tabular}}}} & \multicolumn{1}{c|}{loot}             & \multicolumn{2}{c|}{6.19}           & 6.26                        & \multicolumn{1}{c|}{1.1\%}                        & 5.19                        & \multicolumn{1}{c|}{-16.1\%}                         & 5.18                                    & -16.4\%                                     \\   
\multicolumn{1}{|c|}{}                                                                                & \multicolumn{1}{c|}{red\&black}       & \multicolumn{2}{c|}{9.39}           & 10.20                       & \multicolumn{1}{c|}{8.6\%}                        & 8.15                        & \multicolumn{1}{c|}{-13.2\%}                         & 8.07                                    & -14.1\%                                     \\   
\multicolumn{1}{|c|}{}                                                                                & \multicolumn{1}{c|}{\textbf{average}} & \multicolumn{2}{c|}{7.79}           & 8.23                        & \multicolumn{1}{c|}{4.8\%}                        & 6.67                        & \multicolumn{1}{c|}{\textbf{-14.7\%}}                & 6.62                                    & \textbf{-15.2\%}                            \\ \hline
\multicolumn{1}{|c|}{\multirow{3}{*}{\textbf{\begin{tabular}[c]{@{}c@{}}Owlii\\ vox11\end{tabular}}}} & \multicolumn{1}{c|}{player}           & \multicolumn{2}{c|}{7.72}           & 8.34                        & \multicolumn{1}{c|}{8.0\%}                        & 7.13                        & \multicolumn{1}{c|}{-7.6\%}                          & 6.78                                    & -12.2\%                                     \\   
\multicolumn{1}{|c|}{}                                                                                & \multicolumn{1}{c|}{dancer}           & \multicolumn{2}{c|}{7.80}           & 8.33                        & \multicolumn{1}{c|}{6.9\%}                        & 7.11                        & \multicolumn{1}{c|}{-8.8\%}                          & 6.80                                    & -12.8\%                                     \\   
\multicolumn{1}{|c|}{}                                                                                & \multicolumn{1}{c|}{\textbf{average}} & \multicolumn{2}{c|}{7.76}           & 8.34                        & \multicolumn{1}{c|}{7.4\%}                        & 7.12                        & \multicolumn{1}{c|}{\textbf{-8.2\%}}                 & 6.79                                    & \textbf{-12.5\%}                            \\ \hline
\multicolumn{1}{|c|}{\multirow{3}{*}{\textbf{\begin{tabular}[c]{@{}c@{}}MVUB\\ vox10\end{tabular}}}}  & \multicolumn{1}{c|}{Phil}             & \multicolumn{2}{c|}{10.27}          & 10.13                       & \multicolumn{1}{c|}{-1.4\%}                       & 7.33                        & \multicolumn{1}{c|}{-28.6\%}                         & 6.78                                    & -34.0\%                                     \\   
\multicolumn{1}{|c|}{}                                                                                & \multicolumn{1}{c|}{Ricardo}          & \multicolumn{2}{c|}{5.92}           & 5.12                        & \multicolumn{1}{c|}{-13.6\%}                      & 3.68                        & \multicolumn{1}{c|}{-37.9\%}                         & 3.59                                    & -39.4\%                                     \\   
\multicolumn{1}{|c|}{}                                                                                & \multicolumn{1}{c|}{\textbf{average}} & \multicolumn{2}{c|}{8.10}           & 7.62                        & \multicolumn{1}{c|}{-7.5\%}                       & 5.50                        & \multicolumn{1}{c|}{\textbf{-33.3\%}}                & 5.19                                    & \textbf{-36.7\%}                            \\ \hline
\multicolumn{1}{|c|}{\multirow{2}{*}{\textbf{ScanNet}}}                                               & \multicolumn{1}{c|}{\textbf{q5cm}}    & \multicolumn{2}{c|}{12.92}          & 14.13                       & \multicolumn{1}{c|}{9.3\%}                        & 11.47                       & \multicolumn{1}{c|}{\textbf{-11.2\%}}                & 11.21                                   & \textbf{-13.2\%}                            \\   
\multicolumn{1}{|c|}{}                                                                                & \multicolumn{1}{c|}{\textbf{q2cm}}    & \multicolumn{2}{c|}{13.13}          & 15.04                       & \multicolumn{1}{c|}{14.6\%}                       & 12.04                       & \multicolumn{1}{c|}{\textbf{-8.3\%}}                 & 11.86                                   & \textbf{-9.7\%}                             \\ \hline
\multicolumn{1}{|c|}{\multirow{2}{*}{\textbf{Ford}}}                                                  & \multicolumn{1}{c|}{\textbf{q2cm}}    & \multicolumn{2}{c|}{5.32}           & 7.05                        & \multicolumn{1}{c|}{32.5\%}                       & 5.00                        & \multicolumn{1}{c|}{\textbf{-6.0\%}}                 & -                                       & -                                           \\   
\multicolumn{1}{|c|}{}                                                                                & \multicolumn{1}{c|}{\textbf{q1mm}}    & \multicolumn{2}{c|}{5.22}           & 6.93                        & \multicolumn{1}{c|}{32.9\%}                       & 4.97                        & \multicolumn{1}{c|}{\textbf{-4.7\%}}                 & -                                       & -                                           \\ \hline
\multicolumn{10}{|c|}{\textbf{Average Time (Eocoding Decoding) (s/frame)}}                                                                                                                                                                                                                                                                                                                                                                         \\ \hline
\multicolumn{2}{|c|}{\textbf{8iVFB\_vox10}}                                                                                                   & 9.5    & \multicolumn{1}{c|}{9.3}   & 5.7                         & \multicolumn{1}{c|}{5.1}                          & 10.1                        & \multicolumn{1}{c|}{9.8}                             & 15.7                                    & 16.0                                        \\ 
\multicolumn{2}{|c|}{\textbf{Owlii\_vox11}}                                                                                                   & 32.8   & \multicolumn{1}{c|}{32.0}  & 17.5                        & \multicolumn{1}{c|}{15.3}                         & 37.0                        & \multicolumn{1}{c|}{35.9}                            & 56.0                                    & 58.1                                        \\ 
\multicolumn{2}{|c|}{\textbf{MVUB\_vox10}}                                                                                                    & 17.1   & \multicolumn{1}{c|}{17.0}  & 10.2                        & \multicolumn{1}{c|}{9.2}                          & 19.0                        & \multicolumn{1}{c|}{18.3}                            & 27.3                                    & 28.3                                        \\ 
\multicolumn{2}{|c|}{\textbf{ScanNet\_q2cm}}                                                                                                  & 2.0    & \multicolumn{1}{c|}{2.0}   & 1.2                         & \multicolumn{1}{c|}{1.2}                          & 3.6                         & \multicolumn{1}{c|}{3.5}                             & 6.3                                     & 6.3                                         \\ 
\multicolumn{2}{|c|}{\textbf{Ford\_q1mm}}                                                                                                     & 1.1    & \multicolumn{1}{c|}{1.1}   & 0.8                         & \multicolumn{1}{c|}{0.8}                          & 8.0                         & \multicolumn{1}{c|}{8.0}                             & -                                       & -                                           \\ \hline
\end{tabular}
\end{table}

\subsection{Performance Evaluation}

\textbf{Loss Function.}
We measure the total bits of losslessly compressed attributes as the loss metric in learning, e.g., 
    $Loss = \sum_{s=1}^{S}\sum_{c=1}^{3}\sum_{g=1}^{8}\sum_{i=1}^{N^{(s,g,c)}}-\log_2(p(\hat{x}_{i}^{(s,g,c)}))$
where $s, c$, and $g$ are the scale, color channel, and group index; $N^{(s,g,c)}$ is the number of POVs for color component $c$ in the group $g$ at the scale $s$; and $p$ is calculated using \eqref{eq:entropy}.
\vspace{0.1cm}

{\noindent\bf Datasets.} Extensive experiments are conducted using different datasets shown in Fig.~\ref{fig:dataset} to evaluate the compression efficiency of our method. These datasets include object and scene point clouds from sources with very different geometry and attribute characteristics.

{\it HumanBodies.} 
To measure the compression efficiency on real-life object point clouds, we consider point clouds widely used in standardization committees, including 4 sequences from 8i Voxelized Full Bodies (8iVFB)~\cite{8iVFB}, 4  from Owlii dynamic human mesh (Owlii)~\cite{xu2017owlii}, and 5  from Microsoft Voxelized Upper Bodies (MVUB)~\cite{microsoft2019microsoft}. 
We select 7 of them for training and others for testing, as suggested in~\cite{sheng2021deep,fang20223dac}. The training dataset includes \textit{longdress} \&  \textit{soldier} from 8iVFB, \textit{model} \&  \textit{exercises} from Owlii, and \textit{Andrew}, \textit{David} \&  \textit{Sarah} from MVUB, which are partitioned into 6000 patches for training. Limited by the GPU memory, each patch contains less than 100,000 points. 
The testing dataset includes \textit{loot} \& \textit{redandblack} from 8iVFB, \textit{basketball\_player} \&  \textit{dancer} from Owlii, and \textit{Phil} \&  \textit{Ricardo} from MVUB. All testing samples are prohibited in training.

{\it ScanNet}~\cite{Dai2017ScanNetR3}.
This is a large-scale indoor point cloud dataset totally containing more than 1600 scans, which is widely used in 3D scene understanding tasks. Following common preprocessing methods, raw point clouds are quantized to 2cm and 5cm, respectively. We use 1503 scans for training and 100 scans for the test as suggested.

{\it Ford}~\cite{MPEG_GPCC_CTC}.
Ford is the official outdoor LiDAR dataset used in MPEG, which contains 3 1500-frame sequences. Following the MPEG common test condition, we use the first sequence for training and the other two for testing. Experiments are examined using both original 1mm-precision and quantized 2cm-precision samples.
\vspace{0.1cm}

{\noindent\bf Quantitative Evaluation.} Our test results are shown in Table~\ref{table:main}.
The latest G-PCC version TMC13v14, which provides state-of-the-art performance of lossless PCAC through well-engineered attribute transform~\cite{GPCC}, serves as the anchor. We strictly follow the common test conditions~\cite{MPEG_GPCC_CTC} to generate the anchor results.

As seen, the proposed method offers significant gains against the G-PCC anchor across a variety of datasets:  For object point clouds, we provide 15.2\%, 12.5\%, and 36.7\% bitrate reduction to the G-PCC for 8iVFB, Owlii, and MVUB, respectively; while for indoor/outdoor scene point clouds, we achieve 13.2\% and 9.7\% gains on ScanNet with 5cm and 2cm precision, and 6.0\% and 4.7\% gains on Ford with 2cm and 1mm precision, respectively.

Extensive results suggest that our proposed method generalizes well from dense object point clouds to very sparse LiDAR data, which mainly owes to the joint utilization of cross-scale, cross-group, and cross-color prediction to exploit neighborhood correlations exhaustively. Similar to~\cite{Wang2021SparseTM}, we train the model using two consecutive scales and enforce the model sharing across all scales.

As for the computational complexity measured by the encoding/decoding runtime (seconds per frame or s/frame), our method and the G-PCC share the same order of magnitude when our model is tested on an RTX 3090 GPU, which is attractive for practical applications. Note that these numbers are served as the reference for intuitive understanding because G-PCC and our method run on different platforms (CPU vs. GPU and C++ vs. Python). 

Compared with the well-developed G-PCC, our method only provides a basic learning-based framework that needs further improvement and optimization, especially on sparse point clouds like LiDAR data. 
For example, some advanced technologies in G-PCC~\cite{GPCC} like direct coding mode and angular mode can be borrowed to improve the learned method.

\subsection{Discussion}
\textbf{Modular Impact.} 
We exemplify the performance-complexity tradeoff by examining the impact of  cross-scale, cross-group, and cross-color prediction modules.
As shown in Table~\ref{table:main}, on 8iVFB, the coding gain over the G-PCC increases greatly from the baseline using only cross-scale (CS) prediction with 4.8\% loss, to the case using both cross-scale and cross-group (CG) predictions with 14.7\% compression gain. When the cross-color (CC) prediction is further included, the bitrate reduction  is improved to 15.2\%.  The complexity also increases when involving more prediction modules. In practice, we can carefully choose modules according to the complexity requirement.
\vspace{0.1cm}

{\noindent\bf  Model Generalization.} 
To evidence the generalization of the proposed method, 
we generate a large number of synthetic samples for training~\cite{Wang2022SparsePCAC}. We first densely sample points from raw ShapeNet~\cite{chang2015shapenet} meshes and quantized the coordinates to 8-bit integers. Next, we randomly select images from COCO dataset~\cite{microsoft2014coco} and map them to the points as color attributes through projection, as shown in Fig.~\ref{fig:dataset}. This synthetic dataset provides 6000 object point clouds with different shapes and textures for the training.  
Then we use the models trained on synthesized colorized ShapeNet to test the real-captured object point clouds.   Although the bitrate reduction declines because of very diverse characteristics between training and testing data when comparing with other results in Table~\ref{table:main},  we still achieve averaged 5.2\%, 4.0\%, and  8.8\% positive gain over the G-PCC on 8iVFB, Owlii, and MVUB, respectively,  confidently demonstrating the effectiveness and generalization of our method.
\section{Conclusion}
A lossless PCAC that extensively exploits cross-scale, cross-group and cross-color correlations is developed to accurately approximate attribute probability for high-efficiency compression. The proposed method improves the latest G-PCC with significant bitrate reduction for a variety of diverse point clouds. In the meantime, the computational complexity measured by the encoding/decoding runtime reports the same order of magnitude as that of the G-PCC, which is lightweight and attractive for practical applications. One interesting topic for future exploration is the support of lossy compression under the same framework.

\Section{References}

\bibliographystyle{ieeetr}
\bibliography{pccbib}    

\end{document}